\RequirePackage[2020-02-02]{latexrelease}
\documentclass [aps,amsmath,amssymb,superscriptaddress,preprint]{revtex4}
\usepackage{graphicx}
\usepackage{xcolor}

\newcommand{\lps}{{l_{P}^2}}
\newcommand{\lp}{l_P}

\newcommand{\tp}{t_{P}}

\def\gsim{\mathrel{\rlap{\lower4pt\hbox{\hskip1pt$\sim$}}
		\raise1pt\hbox{$>$}}}       %greater than or approx. symbol

\newcommand{\gn}{G_{\rm N}}

\newcommand{\rs}{R_{\rm s}}

\begin{document}
	\title{Quantum Hair During Gravitational Collapse}
	\author{Xavier Calmet} \email{x.calmet@sussex.ac.uk}
	\affiliation{Department of Physics and Astronomy,\\
		University of Sussex, Brighton, BN1 9QH, United Kingdom}
	\author{Roberto Casadio} \email{casadio@bo.infn.it}
	\affiliation{Dipartimento di Fisica e Astronomia, Universit\`a di Bologna, via Irnerio 46, I-40126 Bologna, Italy}
	\affiliation{
		I.N.F.N., Sezione di Bologna, IS - FLAG, via B.~Pichat~6/2, I-40127 Bologna, Italy}
	\author{Stephen~D.~H.~Hsu} \email{hsusteve@gmail.com}
	\affiliation{Department of Physics and Astronomy\\ Michigan State University \\  East Lansing, Michigan 48823, USA}
	\author{Folkert Kuipers}\email{kuipers@na.infn.it}
	\affiliation{INFN, Sezione di Napoli,\\
		Complesso Universitario di Monte S. Angelo,\\
		Via Cintia Edificio 6, 80126 Napoli, Italy}
	
	\begin{abstract}
		We consider quantum gravitational corrections to the Oppenheimer-Snyder metric describing time-dependent dust ball collapse. The interior metric also describes Friedmann-Lema\^ itre-Robertson-Walker cosmology and our results are interpreted in that context. The exterior corrections are an example of quantum hair, and are shown to persist throughout the collapse. Our results show the quantum hair survives throughout the horizon formation and that the internal state of the resulting black hole is accessible to outside observers.
	\end{abstract}
	
	\maketitle
	\section{Introduction}
	
	The unique quantum gravitational effective action program allows model independent calculations in quantum gravity \cite{Weinberg:1980gg,Barvinsky:1983vpp,Barvinsky:1985an,Barvinsky:1987uw,Barvinsky:1990up,Buchbinder:1992rb,Donoghue:1994dn,Codello:2015mba,Calmet:2018elv}. 
	This approach has been used to study quantum gravitational corrections to a variety of cosmological \cite{Espriu:2005qn,Cabrer:2007xm,Donoghue:2014yha,Codello:2015pga,Calmet:2019tur,Calmet:2016fsr,Calmet:2017hja} and astrophysical models \cite{Calmet:2018rkj,Calmet:2021lny,Calmet:2020tlj,Calmet:2020vuh,Kuipers:2019qby,Calmet:2019eof,Calmet:2017rxl,Calmet:2016sba}. A study of quantum gravitational corrections to a static dust ball used to model a star \cite{Calmet:2019eof,Calmet:2021stu} (see also \cite{Satz:2004hf} for earlier work) revealed the existence of quantum hair. It was found that the quantum gravitational potential of a star depends on the composition of the star at second order in the curvature expansion of the effective action. In  \cite{Calmet:2021stu}, we suggested that the quantum hair would also apply to a collapsing star model and thus also to a black hole.
	
	The aim of the paper is to extend our previous work on quantum hair. We shall first present a very generic result that is independent on the chosen energy-momentum tensor, proving that the quantum hair must exist for any energy-momentum tensor $T_{\mu\nu}$. This result is fully model independent: it does not depend on the matter model (i.e., $T_{\mu\nu}$) or on the high energy completion of the effective action.
	
	We then study a specific model for the gravitational collapse of a dust ball namely the  Oppenheimer-Snyder model of gravitational collapse \cite{Oppenheimer:1939ue} and demonstrate that quantum hair is present in this dynamical model and calculable from first principles. The corrections in $r^{-3}$ and $r^{-5}$ are identical to those identified in \cite{Calmet:2021stu} in the static case. 
	Moreover, the quantum hair persists throughout the gravitational collapse of the star. Our work demonstrates that the resulting black hole has quantum hair.  These results are also relevant to Friedmann-Lema\^ itre-Robertson-Walker (FLRW) cosmology, as the inside of the collapsing object is described by the FLRW metric. We calculate for the first time the complete leading order quantum gravitational correction to FLRW, and comment on previous works on FLRW quantum cosmology.

	This paper is organized as follows. In section 2, we present a model independent proof that quantum hair exists for any energy momentum tensor. In section 3, we review the Oppenheimer-Snyder model. In section 4, we compute the leading quantum gravitational corrections to the interior and exterior metric of this model.  In the conclusions we discuss some of the implications of our work to black hole information and long wavelength quantum gravity.
	
	\section{Quantum hair and generic matter distribution}
	In this section, we use the results presented in \cite{Calmet:2022bpo} and argue that there is quantum hair for any matter distribution. Quantum hair can manifest itself as quantum corrections to classical solutions in general relativity describing the exterior metric of an astrophysical object. In the case of black holes, these quantum corrections can carry information about the interior quantum state, whereas the classical no-hair theorem would forbid this. Hence, the existence of quantum hair bears relevance to the black hole information paradox. 
	
	The quantum corrections to classical solutions of general relativity are reliably calculable using quantum corrected field equations obtained from the variation of the  Vilkovisky-DeWitt unique effective action of quantum gravity as long as curvature invariants remain weak.  At second order in curvature, the effective action is given by
	\begin{equation}
		\Gamma_{\rm QG} = \Gamma_{\rm L} + \Gamma_{\rm NL} \, ,
	\end{equation}
	with a local part 
	\begin{align}
		\Gamma_{\rm L} 
		&= 
		\int d^4x \, \sqrt{|g|} \left[ \frac{M_P^2}{2}
		\big(\mathcal{R} - 2\Lambda\big)
		+ c_1(\mu) \, \mathcal{R}^2 
		+ c_2(\mu) \, \mathcal{R}_{\mu\nu} \mathcal{R}^{\mu\nu} 
		\right.\nonumber\\
		&\qquad \qquad \qquad
		+ c_{3}(\mu) \, \mathcal{R}_{\mu\nu\rho\sigma} \mathcal{R}^{\mu\nu\rho\sigma} 
		+ c_4(\mu) \, \Box \mathcal{R}
		+ \mathcal{O}(M_P^{-2}) \Big],
	\end{align}
	where $M_P=\sqrt{\hbar/ \gn}$ denotes the Planck mass, and a non-local part 
	\begin{align}
		\Gamma_{\rm NL} &= 
		- \int d^4x \, \sqrt{|g|} \left[ 
		\alpha \, \mathcal{R} \ln \left(\frac{\Box}{\mu^2} \right) \mathcal{R}
		+ \beta \, \mathcal{R}_{\mu\nu} \ln \left( \frac{\Box}{\mu^2} \right) \mathcal{R}^{\mu\nu}
		\right.\nonumber\\
		&\qquad \qquad \qquad \qquad \left.
		+ \gamma \, \mathcal{R}_{\mu\nu\rho\sigma} \ln \left( \frac{\Box}{\mu^2} \right) \mathcal{R}^{\mu\nu\rho\sigma}
		+ \mathcal{O}(M_P^{-2}) \right].
	\end{align}
	For simplicity, we set the cosmological constant to zero. In addition, we ignore the boundary term associated with $c_4$, as it does not contribute to the field equations. Then, after applying the local and non-local Gauss-Bonnet identities \cite{Calmet:2018elv}, we obtain
	\begin{align}
		\Gamma_{\rm QG} 
		&=
		\int d^4x \, \sqrt{|g|} \left[ 
		\frac{M_P^2}{2} \,\mathcal{R}
		+ \tilde{c}_1(\mu) \, \mathcal{R}^2 
		+ \tilde{c}_2(\mu) \, \mathcal{R}_{\mu\nu} \mathcal{R}^{\mu\nu} 
		+ \tilde{\alpha} \, \mathcal{R} \ln \left(\frac{\Box}{\mu^2} \right) \mathcal{R}
		\right. \nonumber\\
		& \qquad \qquad \qquad \left.
		+ \tilde{\beta} \, \mathcal{R}_{\mu\nu} \ln \left( \frac{\Box}{\mu^2} \right) \mathcal{R}^{\mu\nu}
		+ \mathcal{O}(M_P^{-2}) \right]
	\end{align}
	with $\tilde{c}_1 = c_1 - c_3$, $\tilde{c}_2 = c_2 + 4 c_3$, $\tilde{\alpha} = \alpha - \gamma$ and $\tilde{\beta} = \beta + 4 \gamma$.

	The quantum gravitational field equations to second order in curvature can be derived from this action, and are given by  
	\begin{eqnarray} \label{FEq}
		{\cal R}_{\mu\nu} - \frac{1}{2}\, {\cal R}\, g_{\mu\nu} - 16\,\pi\,\gn \left( H_{\mu\nu}^{\rm L} + H_{\mu\nu}^{\rm NL} \right)
		= 8 \, \pi \, G_{\rm N} T_{\mu\nu}
		\ ,
	\end{eqnarray}
	where $G_{\rm N}$ is Newton's constant, $T_{\mu\nu}$ is the energy-momentum tensor,
	\begin{align}
		H_{\mu\nu}^{\rm L} 
		=
		&\,  
		\bar{c}_1
		\left( 2\, {\cal R}\, {\cal R}_{\mu\nu} - \frac{1}{2}\, g_{\mu\nu}\, {\cal R}^2 + 2\, g_{\mu\nu}\, \Box {\cal R} - 2 \nabla_\mu \nabla_\nu {\cal R}\right) 
		\label{eq:EQMLoc}
		\\
		&\,
		+\bar{c}_2
		\left( 2\, {\cal R}_{~\mu}^\alpha\, {\cal R}_{\nu\alpha} - \frac{1}{2}\, g_{\mu\nu}\, {\cal R}_{\alpha\beta}\, {\cal R}^{\alpha\beta}
		+ \Box {\cal R}_{\mu\nu} + \frac{1}{2}\, g_{\mu\nu}\, \Box {\cal R}
		- \nabla_\alpha \nabla_\mu {\cal R}_{~\nu}^\alpha
		- \nabla_\alpha \nabla_\nu {\cal R}_{~\mu}^\alpha \right) ,
		\nonumber
	\end{align}
	and
	\begin{align}
		H_{\mu\nu}^{\rm NL} 
		=
		&\,
		- 2\,\alpha
		\left( {\cal R}_{\mu\nu} - \frac{1}{4}\, g_{\mu\nu}\, {\cal R}
		+ g_{\mu\nu}\, \Box
		- \nabla_\mu \nabla_\nu \right)
		\ln\left(\frac{\Box}{\mu^2}\right)\, {\cal R}
		\nonumber
		\\
		&\, 
		- \beta
		\bigg( 2\, \delta_{(\mu}^\alpha\, {\cal R}_{\nu)\beta}
		- \frac{1}{2}\, g_{\mu\nu}\, {\cal R}_{~\beta}^\alpha
		+ \delta_{\mu}^\alpha\, g_{\nu\beta}\, \Box
		+ g_{\mu\nu}\, \nabla^\alpha \nabla_\beta  \nonumber  \\ \nonumber 
		&\quad 
		- \delta_\mu^\alpha\, \nabla_\beta \nabla_\nu
		- \delta_\nu^\alpha\, \nabla_\beta \nabla_\mu 
		\bigg)
		\ln\left(\frac{\Box}{\mu^2}\right)\, {\cal R}_{~\alpha}^\beta &
		\nonumber
		\\
		&\,
		- 2 \,\gamma
		\left(
		\delta_{(\mu}^\alpha\, {\cal R}_{\nu)~\sigma\tau}^{~\beta}
		- \frac{1}{4}\, g_{\mu\nu}\, {\cal R}^{\alpha\beta}_{~~\sigma\tau}
		+\left( \delta_\mu^\alpha\, g_{\nu\sigma} + \delta_\nu^\alpha\, g_{\mu\sigma} \right)
		\nabla^\beta \nabla_\tau \right)
		\ln\left(\frac{\Box}{\mu^2}\right)\, {\cal R}_{\alpha\beta}^{~~\sigma\tau}
		\ .
	\end{align}
	
	With weak curvature invariants, perturbation theory can be applied to solve these complicated coupled partial differential equations, as we can obtain a controlled approximation by expanding in curvature. We thus set $\tilde{g}_{\mu\nu} = g_{\mu\nu} + g_{\mu\nu}^{\rm q}$ where $g_{\mu\nu} $ is the classical solution and  $g_{\mu\nu}^{\rm q}$ the quantum solution one is solving Eq. \eqref{FEq} for. The $\log \Box \mathcal{R}^{\mu...\nu}_{\alpha... \beta}$ terms correspond to kernels that are integrated over curvature terms which are functions of the energy-momentum tensor (see e.g.~Appendix). 
	
	A generic astrophysical body has, relative to its surface or horizon, an interior $T^{\mu\nu}$ tensor and an exterior one. It is clear that the quantum corrections of the outside metric due to the $H_{\mu\nu}^{\rm NL}$ terms must be dependent on $T^{\mu\nu}$ and on the higher curvature terms in the effective action. This fact is independent of the specific type of matter distribution, and thus applies to e.g. a static star, a collapsing star or a real astrophysical black hole (i.e. not a static vacuum solution)\footnote{In the case of a static vacuum solution, quantum hair will still appear at third order in curvature due to higher curvature gravitational terms in the effective action \cite{Calmet:2021lny,Calmet:2017qqa}.}. Hence, the exterior metric will keep a memory of the interior of the matter distribution, which implies the presence of quantum hair for any gravitational body, and, in particular, for realistic black holes. This hair is expressed in terms of deviation from the $1/r$ Newtonian potential. These deviations are due to quantum gravitational corrections to Newton's law.
	
	This is an explicit realisation of the observation that the asymptotic graviton state of an energy eigenstate source is determined at leading
	order by the energy eigenvalue and that the quantum gravitational fluctuations (i.e., graviton loops) produce corrections to the
	long range potential  whose coefficients depend on the internal state of the source \cite{Calmet:2021cip}. We shall consider an explicit application of this result to the Oppenheimer-Snyder gravitational collapse model and calculate the leading order quantum hair to that classical solution.
	
	\section{Oppenheimer-Snyder model: classical Solution}
	We will now consider the Oppenheimer-Snyder model of gravitational collapse \cite{Oppenheimer:1939ue}. In the exterior region, the metric is defined by the line element \cite{Casadio:2010fw}
	\begin{equation}\label{eq:MetricClassicalOutside}
		ds^2 
		= 
		f(R) \, dt^2
		- g(R)^{-1} \, dR^2
		- R^2 \, d \Omega^2 
	\end{equation}
	with
	\begin{equation}\label{eq:frgr}
		f(R) = g(R) = \left( 1 - \frac{2 \, \gn \, M}{R} \right), 
	\end{equation}
	where $M$ is the total ADM mass of the ball, and $R$ is the areal radius  with $R\in[R_s(t),\infty)$. The energy-momentum tensor vanishes: $T_{\mu\nu}=0$. In the interior region, the metric is defined by the line element
	\begin{equation}\label{eq:MetricClassicalInside}
		ds^2 
		= 
		d\tau^2
		- a(\tau)^2 \left( 
		dr^2 + r^2 \, d\Omega^2
		\right)
	\end{equation}
	with $r\in[0,r_s]$, where the scale factor is given by
	\begin{equation}
		a(\tau) = \left(1 - \frac{\tau}{\tau_s} \right)^{2/3},
	\end{equation}
	with $\tau<\tau_s$ and $\tau_s$ the time at which the ball collapses to a singularity. This time can be calculated and is given by
	\begin{equation} \label{taus}
		\tau_s=\sqrt{ \frac{2 \, R_s(0)^3}{9 \,\gn \, M}} \, .
	\end{equation}
	The scale factor corresponds to a Hubble scale
	\begin{equation}\label{eq:HubbleClassical}
		H(\tau) = \frac{\dot{a}(\tau)}{a(\tau)} = \frac{2}{3\,(\tau-\tau_s)} \, .
	\end{equation}
	Furthermore, the energy-momentum tensor is that of a perfect fluid:
	\begin{equation}
		T_{\mu}^{\;\;\nu} = {\rm diag} (-\rho, p, p , p)~~.
	\end{equation}
	The dust ball model assumes $p=w \, \rho$ with $w=0$. 
	%While the dust ball is not a perfect model for a real star (i.e., one cannot completely ignore the pressure throughout the collapse), our focus here is not astrophysics but rather a fundamental issue in quantum gravity. Hence, we can consider a gedanken experiment in which a black hole is formed from collapse of an idealized initial state. We note that as the mass $M$ or radius of the black hole goes to infinity the required density of the progenitor goes to zero: $\rho \sim M / M^3 \sim M^{-2}$ in Planck units. Thus, in a gedanken experiment, one could form a sufficiently massive black hole out of non-interacting objects with very large separation between each of them: for example, small planets, each separated by many times their radius. In this limit interactions are entirely negligible, and the zero pressure dust model applies to very good approximation at large coarse graining scale (e.g., length scales much larger than each planet's radius). 

	%We note that for $r_s \sim \gn \, M$ the dust model is no longer realistic, as the pressure can no longer be ignored. Therefore, for $r_s \lesssim \gn \, M$, the model breaks down and can no longer be trusted as a leading order approximation of collapsing stars.

	\section{Quantum Corrections}
	
	Our goal here is to describe the gravitational collapse of a star, and to show that we can compute, in a controlled approximation, the quantum gravity corrections to the metric during the formation of a black hole. Note that the initial formation of an astrophysical black hole (i.e. non-quantum black hole) does not require large curvatures anywhere in the dust ball. Specifically, this means that we can use the flat space kernel function (see Appendix) to compute quantum corrections arising from the effective action. The only region in the dust ball collapse spacetime with large curvature is the future black hole singularity, but the quantum corrections to the external metric from this region are small as the integrals over the singularity at $r=0$ are well behaved \cite{Calmet:2017qqa,Delgado:2022pcc}.
	
	Thus, we will work again with the Vilkovisky-DeWitt unique effective action of quantum gravity at second order in curvature.
	However, to calculate the quantum correction to the interior metric, it is easiest to use the Weyl basis, in which case one has
	\begin{align}
		\Gamma_{\rm QG} 
		&=
		\int d^4x \, \sqrt{|g|} \left[ 
		\frac{M_P^2}{2} \,\mathcal{R}
		+ \hat{c}_1(\mu) \, \mathcal{R}^2 
		+ \hat{c}_2(\mu) \, \mathcal{C}_{\mu\nu\rho\sigma} \mathcal{C}^{\mu\nu\rho\sigma} 
		+ \hat{\alpha} \, \mathcal{R} \ln \left(\frac{\Box}{\mu^2} \right) \mathcal{R}
		\right. \nonumber\\
		& \qquad \qquad \qquad \left.
		+ \hat{\beta} \, \mathcal{C}_{\mu\nu\rho\sigma} \ln \left( \frac{\Box}{\mu^2} \right) \mathcal{C}^{\mu\nu\rho\sigma}
		+ \mathcal{O}(M_P^{-2}) \right]
	\end{align}
	with $\hat{c}_1 = \tilde{c}_1 + \frac{1}{3} \tilde{c}_2$, $\hat{c}_2 = \frac{1}{2} \tilde{c}_2$, $\hat{\alpha} = \tilde{\alpha} + \frac{1}{3} \tilde{\beta}$ and $\hat{\beta} = \frac{1}{2} \tilde{\beta}$.

	\subsection{Interior corrections}\label{sec:IntCorrections}
	In the interior we have a FLRW metric.  For this metric the Weyl tensor vanishes, and it is thus convenient to work in the Weyl basis. Quantum corrections to this metric have been studied before in the literature. Corrections due to the $\mathcal{R}^2$ term have for example been studied in \cite{Starobinsky:1980te,Capozziello:2014hia,Codello:2015pga}, while non-local corrections due to the $\mathcal{R}\log(\Box)\mathcal{R}$ terms have been studied in \cite{Espriu:2005qn,Cabrer:2007xm,Donoghue:2014yha,Codello:2015mba,Codello:2015pga}. However, none of the previous studies considered the local and non-local corrections together. We will explain that this is crucial for obtaining a consistent result. By considering the local corrections to the action, one obtains the modified Friedmann equation \cite{Codello:2015pga}
	\begin{equation}
		H^2 + \frac{96 \, \pi \, \tp^2 \, c_1(\mu)}{M_P^2} \, \left(2 H \ddot{H} + 6 H^2 \dot{H} - \dot{H}^2  \right) = \frac{8\pi \gn}{3} \, \rho
	\end{equation}
	with $t_P$ the Planck time. We solve this perturbatively, i.e.~we set $H=H_c + H_{\rm L}$, where $H_{\rm L}=\mathcal{O}(\tp^{2})$ and $H_{\rm c}$ solves the classical Friedman equation. It is thus  given by \eqref{eq:HubbleClassical} and
	\begin{equation}
		H_{\rm c}^2 = \frac{8\pi \gn}{3} \, \rho \, .
	\end{equation}
	Solving for $H_{\rm L}$ then yields
	\begin{equation}
		H_{\rm L}(\tau) = \frac{32 \, \pi \, \tp^2 \, \hat{c}_1(\mu)}{(\tau - \tau_s)^3} + \mathcal{O}(\tp^{4}).
	\end{equation}
	
	To obtain the non-local corrections, one must evaluate the $\log \Box \mathcal{R}$ term in the field equations, which is discussed in the Appendix. This leads to the modified Friedmann equation \cite{Codello:2015pga} given by
	\begin{equation}
		H^2(\tau) - \frac{256\,\pi\, \tp^2\, \hat{\alpha}}{3\,(\tau_s-\tau)^4} \left[\ln(\mu\, \tau) + \ln\left(1 - \frac{\tau}{\tau_s} \right) - \frac{2 \, \tau}{3 \, \tau_s} \right] = \frac{8\pi \gn}{3} \rho(\tau).
	\end{equation}
	Solving this perturbatively yields
	\begin{equation}
		H_{\rm NL}(t) = \frac{64 \, \pi \, \tp^2 \, \hat{\alpha}}{(\tau-\tau_s)^3} \left[ - \frac{2\,\tau}{3\, \tau_s} + \ln\left(1 - \frac{\tau}{\tau_s} \right) + \ln(\mu \,\tau) \right]+ \mathcal{O}(\tp^{4}).
	\end{equation}
	Gathering the local and non-local corrections, we find
	\begin{equation}
		H(\tau) 
		= 
		\frac{2}{3\,(\tau-\tau_s)} \left\{ 
		1 
		+ \frac{48 \, \pi \, \tp^2}{(\tau - \tau_s)^2} \left[ 
		\hat{c}_1(\mu) 
		+ 2 \, \hat{\alpha} \left( 
		\ln\left[
		\mu \, \tau \left(
		1 - \frac{\tau}{\tau_s} 
		\right) 
		\right]
		- \frac{2 \, \tau}{3 \, \tau_s}
		\right) 
		\right] 
		+ \mathcal{O}(\tp^{4}) 
		\right\}.
	\end{equation}
	This equation is renormalization group invariant because we have considered the local and non-local corrections together. This had not been done previously in the literature, implying that previous studies of FLRW cosmology within this framework are flawed. 
	We can now solve for $a(\tau)$. We find
	\begin{align}
		a(\tau)
		&= 
		\exp\left(\int_0^\tau H(s) \, ds \right)\nonumber\\
		&=
		\left(1 - \frac{\tau}{\tau_s}  \right)^{2/3}
		\left(1 + \frac{16 \, \pi \, \tp^2}{(\tau_s - \tau)^2} \left\{
		\hat{c}_1(\mu)\, \frac{\tau}{\tau_s} \left(\frac{\tau}{\tau_s} - 2  \right)
		\right. \right. \nonumber\\
		&\quad \left. \left.
		+ 2\, \hat{\alpha}  \, \left[ 
		\frac{\tau^2}{6\, \tau_s^2} 
		+ \frac{\tau}{\tau_s} \left(\frac{\tau}{\tau_s} - 2  \right) \ln(\mu\,\tau)
		- \left(\frac{\tau^2}{\tau_s^2} - \frac{2\,\tau}{\tau_s} + 2  \right) \ln\left(1 - \frac{\tau}{\tau_s}  \right)
		\right]
		\right\}
		+ \mathcal{O}(\tp^4) \right),
	\end{align}
	where $\tau\in[0,\tau_s)$. This is the quantum correction to the interior metric. Note that it depends on both the non-local and the local Wilson coefficients. The latter are not calculable within the effective theory approach that we have used, as it depends on the ultra-violet completion of the effective action. 
	
	Because we have considered the full effective action to second order in curvature, our result differs from previous studies of quantum cosmology within this framework, such as that in \cite{Donoghue:2014yha} where only the non-local contributions were considered. The phenomenology clearly needs to be considered again and the question of a big bounce should be investigated anew.  We now turn our attention to the exterior solution and its quantum gravitational corrections.

	\subsection{Exterior corrections}\label{sec:ExtCorrections}
	We discuss the kernel in spherically symmetric coordinates in the Appendix. The calculation follows the methodology of that presented in \cite{Calmet:2019eof} with the notable complication that we now have a space-time dependent problem. In the Appendix we show that $\ln(\Box)\mathcal{R}$ can be approximated by eq.~\eqref{eq:KernelExterior}
	\begin{equation*}
		\ln\left(\frac{\Box}{\mu^2} \right) \mathcal{R}(x)
		=
		- \frac{2 \, G_{\rm N} \, M}{3 \, R_s(0)^3} \left\{\frac{2\, R_s(t_r)}{r} + \ln\left[\frac{r - R_s(t_r)}{r + R_s(t_r)} \right] \right\}
		+ {\cal O}(\dot{R}_s),
	\end{equation*}
	where
	\begin{equation}
		\dot{R}_s(t_r) = \frac{d R_s(t_r)}{dt_r}
	\end{equation}
	denotes a derivative with respect to the retarded time coordinate as measured by a distant observer. As this derivative remains small during the formation of the black hole at $t\rightarrow\infty$, the corrections encapsulated in the term ${\cal O}(\dot{R}_s)$ remain subleading throughout the collapse. 
	We note that, at this order in the derivative expansion, one can easily obtain similar expressions for  the corrections due to $\ln(\Box)\mathcal{R}_{\mu\nu}$ and $\ln(\Box)\mathcal{C}_{\mu\nu\rho\sigma}$, cf. e.g. \cite{Calmet:2019eof}.
	\par 
	
	Using these results, the quantum corrected Einstein equations \eqref{FEq} can be solved perturbatively. This yields a correction to the functions $f(R)$ and $g(R)$ defined in eq.~\eqref{eq:frgr}. At leading order these corrections are given by
	\begin{align}
		\delta f(t_r,R)
		&=
		(\alpha + \beta + 3\, \gamma) \, \frac{192\, \pi \, \lps \, G_{\rm N} \, M}{R_s(0)^3} \left\{ \frac{2 \, \rs(t_r)}{R}  + \ln\! \left[ \frac{R-\rs(t_r)}{R+R_s(t_r)} \right] \right\} + {\cal O}(\dot{R}_s) ,
		\\
		\delta g(t_r,R)
		&=
		(\alpha - \gamma) \, \frac{384 \, \pi \, \lps \, G_{\rm N} \, M }{R_s(0)^3} 
		\frac{R_s(t_r)^3}{R \, [R^2 - R_s(t_r)^2]} + {\cal O}(\dot{R}_s)
	\end{align}
	which coincides with the results for the static dust ball given in Ref.~\cite{Calmet:2021stu} after making the replacement $R_s \to R_s(t_r)$. In the above $l_P=\sqrt{\hbar \gn}$ is the Planck length.
	\par 
	
	%At this order in the derivative expansion the corrections only depend on the density distribution at $t_r$. Corrections appearing at higher order in this expansion integrate over the trajectory from the start of the collapse up to $t_r$. Hence, the presence of higher derivative terms implies a memory effect, as the corrections depend on the collapse trajectory. This memory effect is suppressed in the early stage of the collapse where $\dot{R}_s\ll 1$, but becomes relevant at late times, when $\dot{R}_s\rightarrow 1$.
	%\par
	
	Let us focus on the $tt$-component of the metric and expand the result for $R\gg R_s$ making the different expansion parameters explicit. We obtain
	\begin{eqnarray}
		f(t_r,R) 
		&=&1
		-\frac{2 \, G_{\rm N} \, M}{R}
		- 128 \, \pi^2 \, (\alpha + \beta + 3\, \gamma) \, \frac{\lps}{R^2} \left\{ \frac{G_{\rm N} \, M}{R} \, \frac{R_s(t_r)^3}{R_s(0)^3}
		\right.\nonumber\\
		&&\qquad \left. \times
		\left[ 1 + \frac{3 \, R_s(t_r)^2}{5 \, R^2} + {\cal O} \left(\frac{R_s(t_r)}{R}\right)^4\right] + {\cal O}(\dot{R}_s) \right\}
		+ \mathcal{O}\left(\frac{\lp}{R}\right)^4 \ .
	\end{eqnarray}
	The expansion in $l_P/R$ reflects the truncation of the effective action at second order in curvature.
	\par 
	
	Taking into account these expansion parameters, we can state the main result as follows: we have computed the coefficient of a (fully quantum mechanical) correction to the exterior metric which behaves as $R^{-5}$ asymptotically far from the black hole. At leading order this coefficient depends on the density distribution of the dust ball from which the black hole was formed. Corrections to this result are suppressed by factors of order $l_P / R_s$ and $\dot{R_s}$.
	
	Our result implies that the quantum hair identified in \cite{Calmet:2021stu} is present in the collapse (i.e., space-time dependent) model considered here. This correction survives throughout the gravitational collapse. Our result is  further evidence that black holes have quantum hair.

	\section{Conclusions}
	In this paper, we have revisited the question of quantum hair and quantum memory in quantum gravity. Within the context of the unique effective action, we have shown that quantum hair and quantum memory are very generic features of quantum gravity and that any non-zero energy momentum tensor will produce quantum hair in the form of quantum corrections to the classical spacetime resulting from the Einstein equations. While the general proof is model independent, we have illustrated this result with a direct calculation in the case of the Oppenheimer-Snyder collapse model. We have shown by explicit calculation that the quantum corrections to the Oppenheimer-Snyder classical solution are sensitive to the density of the matter distribution. The outside gravitational field contains information about the collapse process that is stored in the quantum hair. In principle, a distant observer could measure the deviation from the Newton potential. This work is a further demonstration that all classical solutions in general relativity, including black holes, are hairy in quantum gravity.
	
	\bigskip
	
	{\it Acknowledgments:}
	The work of X.C. is supported in part  by the Science and Technology Facilities Council (grants numbers ST/T00102X/1 and ST/T006048/1). The work of R.C. is partially supported by the INFN grant FLAG, it has also been carried out in the framework of activities of the National Group of Mathematical Physics (GNFM, INdAM). The research of F.K. was carried out in the frame of Programme STAR Plus, financially supported  by UniNA and Compagnia di San Paolo. 
	\\
	\bigskip 
	
	{\it Data Availability Statement:}
	This manuscript has no associated data. Data sharing not applicable to this article as no datasets were generated or analysed during the current study.
	
	\bigskip 
	
	%\newpage

	\section*{Appendix} 
	We are interested in evaluating the expression
	\begin{equation}\label{LogBoxR}
		\ln\left(\frac{\Box}{\mu^2} \right) \mathcal{R}(x) \, ,
	\end{equation}
	which we can write as
	\begin{equation}\label{eq:BoxIntegralExpres}
		\int d^4x' \sqrt{|g(x')|} \, L(x-x') \, \mathcal{R}(x') \, .
	\end{equation}
	As we are working at second order in curvature, we will approximate the kernel $L$ by its flat space kernel $L^{\rm flat}$ that is given in eq.~(55) of Ref.~\cite{Calmet:2018rkj}
	\begin{align}\label{eq:FlatKernel}
		L^{\rm flat}(x-x') 
		&= 
		\lim_{\delta \rightarrow 0} \left[
		\frac{i}{\pi^2} \left(
		\frac{\Theta(\Delta t ) \, \Theta\left((x-x')^2\right)  }{\left[(\Delta t + i \, \delta )^2 - (\vec{x} - \vec{x}' )^2 \right]^2}
		- 	\frac{\Theta(\Delta t ) \, \Theta\left((x-x')^2\right) }{\left[(\Delta t - i \, \delta )^2 - (\vec{x} - \vec{x}' )^2 \right]^2}
		\right) 
		\right.
		\nonumber\\
		&\qquad \quad
		- 2\, \delta^{(4)}(x-x') \, \ln(\delta \, \mu) 
		\Big]
	\end{align}
	with $\Delta t := t-t'$. We will be interested in evaluating this kernel in two regimes:
	\begin{itemize}
		\item the interior of the collapsing star described by the FLRW metric \eqref{eq:MetricClassicalInside}; 
		\item the exterior of the collapsing star described by the Schwarschild metric \eqref{eq:MetricClassicalOutside}.
	\end{itemize}

	\subsection{Interior}
	In the interior coordinate system $(\tau,r,\theta,\phi)$, the Ricci scalar is given by
	\begin{equation}\label{eq:RicciS}
		\mathcal{R}(x) = -\frac{4 \, \Theta(r_s - r)}{3\, \tau_s^2 \, a(\tau)^3} \, .
	\end{equation}
	Using this expression, we can perform the spatial integrals in eq.~\eqref{eq:BoxIntegralExpres}. This yields
	\begin{align}\label{LogBoxInterior}
		\ln\left(\frac{\Box}{\mu^2} \right) \mathcal{R}(x)
		&=
		\frac{4}{3 \, \tau_s^2} \, \int_{\tau-(r_s+r)}^{\tau-(r_s-r)} \left( \frac{1}{\tau - \tau'}    - \frac{1}{r} \right) d\tau'
		\nonumber\\
		&\quad
		+ \frac{8}{3 \, \tau_s^2}   \lim _{\delta \rightarrow 0} \left( \int_{t-(r_s-r)}^{t-\delta} \frac{1}{\tau-\tau'} \,   d\tau' 
		+ \ln(\delta \mu) \right),
	\end{align}
	which reduces to the results obtained in Refs.~\cite{Donoghue:2014yha,Codello:2015pga} in the limit $r_s\rightarrow\infty$.
	In deriving this result we have approximated the function $\ln(\Box)$ by its kernel in flat spacetime \eqref{eq:FlatKernel}, which is a valid approximation at second order in curvature.
	
	\subsection{Exterior}
	In the exterior coordinate system $(t,R,\theta,\phi)$, the Ricci scalar \eqref{eq:RicciS} can be written as
	\begin{equation}
		\mathcal{R}(x) = -\frac{6 \, G_{\rm N} \, M \, \Theta(R_s(t) - R)}{R_s(0)^3 \, a(t)^3}  \, .
	\end{equation}
	Using this expression, we can perform the spatial integrals in eq.~\eqref{eq:BoxIntegralExpres}. This yields
	\begin{align}\label{LogBoxExterior}
		\ln\left(\frac{\Box}{\mu^2} \right) \mathcal{R}(x)
		&=
		\frac{6 \, G_{\rm N} \, M}{R_s(0)^3} \, \int_{-\infty}^{t} \Theta[R_s(t') + t' - t + R] \, \Theta[R_s(t') - t' + t - R] \left( \frac{1}{t - t'}    - \frac{1}{R} \right) dt' \, .
	\end{align}
	As was the case for the interior calculation, this result relies on approximating the function $\ln(\Box)$ by its flat space kernel \eqref{eq:FlatKernel}, which is valid in the coordinate frame of a distant observer, where the collapse remains slow. Corrections to the result appear at ${\cal O}(\dot{R}_s^2,\ddot{R}_s R_s)$, where the derivatives are taken with respect to the retarded time coordinate $t_r$.
	\par 
	
	We can further evaluate the time integral in eq.~\eqref{LogBoxExterior}. The domain of 
	integration of this expression in $t'$ is determined by the two Heaviside functions, which require
	\begin{equation}\label{intdomain}
		t - R - R_s(t') \leq t' \leq t - R + R_s(t') \, .
	\end{equation}
	We note that $R_s(t)=r_s\, a(\tau)$, where $\tau=\tau(t)$ and $r_s=R_s(0)$. However, at leading order in the derivative expansion the task becomes much simpler.
	In general, the integration domain \eqref{intdomain} will be given by
	\begin{equation}
		T_-(t_r;R_s(t_r),\dot{R}_s(t_r))
		\le
		t'
		\le
		T_+(t_r;R_s(t_r),\dot{R}_s(t_r))
		<
		t
		\ ,
	\end{equation}
	where we explicitly showed that the endpoints $T_\pm$ depend on both the retarded time $t_r$ and the radius of the star evaluated at the retarded time $R_s(t_r)$. We note that the integration bounds may also depend on higher derivatives, but these terms are suppressed, as they appear at higher order in the derivative expansion.
	\par
	
	At this order in the derivative expansion, we can Taylor expand $R_s(t')$, which yields
	\begin{equation}
		R_s(t') = R_s(t_r) + \dot{R}_s(t_r)\, (t' - t_r) + {\cal O}(\ddot{R}_s)
		\ .
	\end{equation}
	Using this expansion, we find that the range of integration in $t'$ is explicitly determined by
	\begin{equation}
		T_-
		=
		\frac{t_r - R_s(t_r) + t_r \, \dot{R}_s(t_r)}{1 +\dot{R}_s(t_r)}
		\lesssim 
		t'
		\lesssim
		\frac{t_r + R_s(t_r) - t_r \, \dot{R}_s(t_r)}{1 - \dot{R}_s(t_r)}
		=
		T_+
		\ .
		\label{dt'full}
	\end{equation}
	Hence, up to first order time derivatives, we obtain
	\begin{equation}
		t_r - R_s(t_r) [1 - \dot{R}_s(t_r)]
		\lesssim 
		t'
		\lesssim
		t_r + R_s(t_r) [1 + \dot{R}_s(t_r)]
		\ .
		\label{dt'}
	\end{equation}

	Therefore, eq.~\eqref{LogBoxExterior} simplifies to
	\begin{align}
		\ln\left(\frac{\Box}{\mu^2} \right) \mathcal{R}(x)
		&=
		\frac{6 \, G_{\rm N} \, M}{R_s(0)^3} \, \int_{t_r - R_s(t_r) [1 - \dot{R}_s(t_r)]}^{t_r + R_s(t_r) [1 + \dot{R}_s(t_r)] } \left( \frac{1}{t - t'}    - \frac{1}{R} \right) dt' \,  .
	\end{align}
	Evaluating this integral, we find our result
	\begin{align}
		\ln\left(\frac{\Box}{\mu^2} \right) \mathcal{R}(x)
		&=
		- \frac{6 \, G_{\rm N} \, M}{R_s(0)^3} \left\{\frac{2\, R_s(t_r)}{R} + \ln\left[\frac{R - R_s(t_r)[1 + \dot{R}_s(t_r)]}{R + R_s(t_r)[1 - \dot{R}_s(t_r)]} \right] \right\} + {\cal O}(\dot{R}_s^2, \ddot{R}_s \, R_s) \, ,
	\end{align}
	which reduces to the result obtained in Ref.~\cite{Calmet:2019eof} for constant radius $R_s(t_r)=R_s$. We can further approximate this result by
	\begin{align}\label{eq:KernelExterior}
		\ln\left(\frac{\Box}{\mu^2} \right) \mathcal{R}(x)
		&=
		- \frac{6 \, G_{\rm N} \, M}{R_s(0)^3} \left\{\frac{2\, R_s(t_r)}{R} + \ln\left[\frac{R - R_s(t_r)}{R + R_s(t_r)} \right] \right\} + {\cal O}(\dot{R}_s) \, ,
	\end{align}
	which is the result applied in section \ref{sec:ExtCorrections}.
	\par 
	
	Let us emphasize that all expressions hold throughout the formation of a black hole at $t\rightarrow\infty$. In this regime $|\dot{R}_s|\lesssim\frac{1}{6}$, implying that the perturbative expansion is under control. Further corrections can be calculated reliably, but the calculations are rather complicated and the results difficult to display.


\begin{thebibliography}{10}
		%\cite{Weinberg:1980gg}
		\bibitem{Weinberg:1980gg} 
		S.~Weinberg,
		%``Ultraviolet Divergences In Quantum Theories Of Gravitation,''
		in \textit{General Relativity: An Einstein Centenary Survey}, Cambridge, UK, 790 (1980).
		
		%\cite{Barvinsky:1983vpp}
		\bibitem{Barvinsky:1983vpp}
		A.~O.~Barvinsky and G.~A.~Vilkovisky,
		%``The Generalized Schwinger-DeWitt Technique and the Unique Effective Action in Quantum Gravity,''
		Phys. Lett. B \textbf{131}, 313-318 (1983).
		%doi:10.1016/0370-2693(83)90506-3
		
		%\cite{Barvinsky:1985an}
		\bibitem{Barvinsky:1985an} 
		A.~O.~Barvinsky and G.~A.~Vilkovisky,
		%``The Generalized Schwinger-Dewitt Technique in Gauge Theories and Quantum Gravity,''
		Phys.\ Rept.\  {\bf 119}, 1 (1985).
		%doi:10.1016/0370-1573(85)90148-6
		
		%\cite{Barvinsky:1987uw}
		\bibitem{Barvinsky:1987uw} 
		A.~O.~Barvinsky and G.~A.~Vilkovisky,
		%``Beyond the Schwinger-Dewitt Technique: Converting Loops Into Trees and In-In Currents,''
		Nucl.\ Phys.\ B {\bf 282}, 163 (1987).
		%doi:10.1016/0550-3213(87)90681-X
		
		%\cite{Barvinsky:1990up}
		\bibitem{Barvinsky:1990up} 
		A.~O.~Barvinsky and G.~A.~Vilkovisky,
		%``Covariant perturbation theory. 2: Second order in the curvature. General algorithms,''
		Nucl.\ Phys.\ B {\bf 333}, 471 (1990).
		%doi:10.1016/0550-3213(90)90047-H
		
		%  \cite{Buchbinder:1992rb}
		\bibitem{Buchbinder:1992rb} 
		I.~L.~Buchbinder, S.~D.~Odintsov and I.~L.~Shapiro,
		``Effective action in quantum gravity,''
		(CRC Press, Bristol, 1992)
		
		%\cite{Donoghue:1994dn}
		\bibitem{Donoghue:1994dn} 
		J.~F.~Donoghue,
		%``General relativity as an effective field theory: The leading quantum corrections,''
		Phys.\ Rev.\ D {\bf 50}, 3874 (1994)
		%doi:10.1103/PhysRevD.50.3874
		%[gr-qc/9405057].
		
		\bibitem{Codello:2015mba}
		A.~Codello and R.~K.~Jain,
		``On the covariant formalism of the effective field theory of gravity and leading order corrections,''
		Class. Quant. Grav. \textbf{33}, no.22, 225006 (2016).
		%doi:10.1088/0264-9381/33/22/225006
		%[arXiv:1507.06308 [gr-qc]].
		
		
		%\cite{Calmet:2018elv}
		\bibitem{Calmet:2018elv}
		X.~Calmet,
		%``Vanishing of Quantum Gravitational Corrections to Vacuum Solutions of General Relativity at Second Order in Curvature,''
		Phys. Lett. B \textbf{787} (2018), 36-38.
		%doi:10.1016/j.physletb.2018.10.040
		%[arXiv:1810.09719 [hep-th]].
		
		\bibitem{Donoghue:2014yha}
		J.~F.~Donoghue and B.~K.~El-Menoufi,
		``Nonlocal quantum effects in cosmology: Quantum memory, nonlocal FLRW equations, and singularity avoidance,''
		Phys. Rev. D \textbf{89}, no.10, 104062 (2014).
		%doi:10.1103/PhysRevD.89.104062
		%[arXiv:1402.3252 [gr-qc]].
		
		\bibitem{Codello:2015pga}
		A.~Codello and R.~K.~Jain,
		``On the covariant formalism of the effective field theory of gravity and its cosmological implications,''
		Class. Quant. Grav. \textbf{34}, no.3, 035015 (2017).
		%doi:10.1088/1361-6382/aa549d
		%[arXiv:1507.07829 [astro-ph.CO]].
		
		
		
		
		
		
		\bibitem{Espriu:2005qn}
		D.~Espriu, T.~Multamaki and E.~C.~Vagenas,
		``Cosmological significance of one-loop effective gravity,''
		Phys. Lett. B \textbf{628}, 197-205 (2005).
		%doi:10.1016/j.physletb.2005.09.033
		%[arXiv:gr-qc/0503033 [gr-qc]].
		
		\bibitem{Cabrer:2007xm}
		J.~A.~Cabrer and D.~Espriu,
		``Secular effects on inflation from one-loop quantum gravity,''
		Phys. Lett. B \textbf{663}, 361-366 (2008)
		%doi:10.1016/j.physletb.2008.04.047
		%[arXiv:0710.0855 [gr-qc]].
		
		
		%\cite{Calmet:2019tur}
		\bibitem{Calmet:2019tur}
		X.~Calmet, J.~Edholm and I.~Kuntz,
		%``Imprints of Quantum Gravity in the Cosmic Microwave Background,''
		Eur. Phys. J. C \textbf{79} (2019) no.3, 238
		doi:10.1140/epjc/s10052-019-6756-x
		[arXiv:1903.01379 [hep-th]].
		
		%\cite{Calmet:2016fsr}
		\bibitem{Calmet:2016fsr}
		X.~Calmet and I.~Kuntz,
		%``Higgs Starobinsky Inflation,''
		Eur. Phys. J. C \textbf{76} (2016) no.5, 289
		doi:10.1140/epjc/s10052-016-4136-3
		[arXiv:1605.02236 [hep-th]].
		
		%\cite{Calmet:2017hja}
		\bibitem{Calmet:2017hja}
		X.~Calmet, I.~Kuntz and I.~G.~Moss,
		%``Non-Minimal Coupling of the Higgs Boson to Curvature in an Inflationary Universe,''
		Found. Phys. \textbf{48} (2018) no.1, 110-120
		doi:10.1007/s10701-017-0131-2
		[arXiv:1701.02140 [hep-ph]].
		
		
		
		
		
		%\cite{Calmet:2018rkj}
		\bibitem{Calmet:2018rkj}
		X.~Calmet, B.~K.~El-Menoufi, B.~Latosh and S.~Mohapatra,
		%``Gravitational Radiation in Quantum Gravity,''
		Eur. Phys. J. C \textbf{78} (2018) no.9, 780
		doi:10.1140/epjc/s10052-018-6265-3
		[arXiv:1809.07606 [hep-th]].
		
		%\cite{Calmet:2021lny}
		\bibitem{Calmet:2021lny}
		X.~Calmet and F.~Kuipers,
		%``Quantum gravitational corrections to the entropy of a Schwarzschild black hole,''
		Phys. Rev. D \textbf{104} (2021) no.6, 066012
		doi:10.1103/PhysRevD.104.066012
		[arXiv:2108.06824 [hep-th]].
		
		%\cite{Calmet:2020tlj}
		\bibitem{Calmet:2020tlj}
		X.~Calmet, R.~Casadio and F.~Kuipers,
		%``Quantum corrected equations of motion in the interior and exterior Schwarzschild spacetime,''
		Phys. Rev. D \textbf{102} (2020) no.2, 026018
		doi:10.1103/PhysRevD.102.026018
		[arXiv:2007.05416 [hep-th]].
		
		%\cite{Calmet:2020vuh}
		\bibitem{Calmet:2020vuh}
		X.~Calmet, R.~Casadio and F.~Kuipers,
		%``Singularities in quantum corrected space-times,''
		Phys. Lett. B \textbf{807} (2020), 135605
		doi:10.1016/j.physletb.2020.135605
		[arXiv:2003.04220 [hep-th]].
		
		%\cite{Kuipers:2019qby}
		\bibitem{Kuipers:2019qby}
		F.~Kuipers and X.~Calmet,
		%``Singularity theorems in the effective field theory for quantum gravity at second order in curvature,''
		Universe \textbf{6} (2020) no.10, 171
		doi:10.3390/universe6100171
		[arXiv:1911.05571 [gr-qc]].
		
		%\cite{Calmet:2019eof}
		\bibitem{Calmet:2019eof}
		X.~Calmet, R.~Casadio and F.~Kuipers,
		%``Quantum Gravitational Corrections to a Star Metric and the Black Hole Limit,''
		Phys. Rev. D \textbf{100} (2019) no.8, 086010
		doi:10.1103/PhysRevD.100.086010
		[arXiv:1909.13277 [hep-th]].
		
		%\cite{Calmet:2017rxl}
		\bibitem{Calmet:2017rxl}
		X.~Calmet, S.~Capozziello and D.~Pryer,
		%``Gravitational Effective Action at Second Order in Curvature and Gravitational Waves,''
		Eur. Phys. J. C \textbf{77} (2017) no.9, 589
		doi:10.1140/epjc/s10052-017-5172-3
		[arXiv:1708.08253 [hep-th]].
		
		%\cite{Calmet:2016sba}
		\bibitem{Calmet:2016sba}
		X.~Calmet, I.~Kuntz and S.~Mohapatra,
		%``Gravitational Waves in Effective Quantum Gravity,''
		Eur. Phys. J. C \textbf{76} (2016) no.8, 425
		doi:10.1140/epjc/s10052-016-4265-8
		[arXiv:1607.02773 [hep-th]].
		
		
		
		
		%\cite{Calmet:2021stu}
		\bibitem{Calmet:2021stu}
		X.~Calmet, R.~Casadio, S.~D.~H.~Hsu and F.~Kuipers,
		%``Quantum Hair from Gravity,''
		Phys. Rev. Lett. \textbf{128}, no.11, 111301 (2022)
		doi:10.1103/PhysRevLett.128.111301
		[arXiv:2110.09386 [hep-th]].
		%5 citations counted in INSPIRE as of 05 Apr 2022
		
		
		%\cite{Satz:2004hf}
		\bibitem{Satz:2004hf}
		A.~Satz, F.~D.~Mazzitelli and E.~Alvarez,
		%``Vacuum polarization around stars: Nonlocal approximation,''
		Phys. Rev. D \textbf{71} (2005), 064001
		doi:10.1103/PhysRevD.71.064001
		[arXiv:gr-qc/0411046 [gr-qc]].
		
		%\cite{Calmet:2021cip}
		\bibitem{Calmet:2021cip}
		X.~Calmet and S.~D.~H.~Hsu,
		%``Quantum hair and black hole information,''
		Phys. Lett. B \textbf{827} (2022), 136995
		doi:10.1016/j.physletb.2022.136995
		[arXiv:2112.05171 [hep-th]].
		
		%\cite{Calmet:2022bpo}
		\bibitem{Calmet:2022bpo}
		X.~Calmet and S.~D.~H.~Hsu,
		%``Quantum Hair in Electrodynamics and Gravity,''
		[arXiv:2209.12798 [hep-th]].
		
		\bibitem{Oppenheimer:1939ue}
		J.~R.~Oppenheimer and H.~Snyder,
		``On Continued gravitational contraction,''
		Phys. Rev. \textbf{56}, 455-459 (1939).
		%doi:10.1103/PhysRev.56.455
		
		\bibitem{Casadio:2010fw}
		R.~Casadio, S.~D.~H.~Hsu and B.~Mirza,
		``Asymptotic Safety, Singularities, and Gravitational Collapse,''
		Phys. Lett. B \textbf{695}, 317-319 (2011).
		%doi:10.1016/j.physletb.2010.10.060
		%[arXiv:1008.2768 [gr-qc]].
		
		
		
		
		%\cite{Starobinsky:1980te}
		\bibitem{Starobinsky:1980te}
		A.~A.~Starobinsky,
		%``A New Type of Isotropic Cosmological Models Without Singularity,''
		Phys. Lett. B \textbf{91} (1980), 99-102
		doi:10.1016/0370-2693(80)90670-X
		
		%\cite{Capozziello:2014hia}
		\bibitem{Capozziello:2014hia}
		S.~Capozziello, M.~De Laurentis and O.~Luongo,
		%``Connecting early and late universe by $f(R)$ gravity,''
		Int. J. Mod. Phys. D \textbf{24} (2014) no.04, 1541002
		doi:10.1142/S0218271815410023
		[arXiv:1411.2822 [gr-qc]].
		%\cite{Calmet:2017qqa}
		\bibitem{Calmet:2017qqa}
		X.~Calmet and B.~K.~El-Menoufi,
		%``Quantum Corrections to Schwarzschild Black Hole,''
		Eur. Phys. J. C \textbf{77} (2017) no.4, 243.
		%doi:10.1140/epjc/s10052-017-4802-0
		%[arXiv:1704.00261 [hep-th]].
		%29 citations counted in INSPIRE as of 27 Sep 2021
		
		\bibitem{Delgado:2022pcc}
		R.~C.~Delgado,
		%``Quantum gravitational corrections to the entropy of a Reissner\textendash{}Nordstr\"om black hole,''
		Eur. Phys. J. C \textbf{82} (2022) no.3, 272
		doi:10.1140/epjc/s10052-022-10232-0
		[arXiv:2201.08293 [hep-th]].
		
		
		
		
		
		
		
		
		
		
		
		
		
		
		
	\end{thebibliography}
\end{document}